\documentclass[%
 aip,
cp,  
 amsmath,amssymb,
 reprint,%
]{revtex4-2}

\usepackage{graphicx}
\usepackage{dcolumn}
\usepackage{bm}

\usepackage[utf8]{inputenc}
\usepackage[T1]{fontenc}
\usepackage{mathptmx} 
\usepackage{float}
\usepackage{hyperref}

\begin{document}

\title{Scientific Computing with Open SageMath \\
not only for Physics Education
}

\author{Dominik Borovsk\'y}
 \email{dominik.borovsky@student.upjs.sk}
\author{Jozef Han\v{c}} 
 \email[\textit{corresponding author:} ]{jozef.hanc@upjs.sk}
\affiliation{
  Institute of Physics, Faculty of Science, P. J. Šafárik University in Košice, Slovakia 
}
\author{Martina Han\v{c}ov\'a}
 \email{martina.hancova@upjs.sk}
\affiliation{%
Institute of Mathematics, Faculty of Science, P. J. Šafárik University in Košice, Slovakia 
}%


\date{\today} 

\begin{abstract}
Nowadays interactive digital scientific environments have become an integral part of scientific computing in solving various scientific tasks in research, but also STEM education. We introduce SageMath or shortly Sage -- a free open Python-based alternative to the well-known commercial software -- in the frame of our course Methods of Physical Problems Solving for future scientists and science teachers. Particularly, in the 1st illustrative example from the Physics Olympiad, we present Sage as a scientific open data source, symbolic, numerical, and visualization tool. The 2nd example from the Young Physicists' Tournament shows Sage as a multimedia, modeling, and programming tool. By employing SageMath as an open digital environment for scientific computing in the education of all STEM disciplines, teachers and students are empowered not only with a universal educational tool, but a real research tool, enabling them to engage in interactive visualization, modeling, programming, and solving of authentic, complex interdisciplinary problems, thus naturally enhancing their motivation to pursue science in alignment with the core mission of STEM education.
\end{abstract}

\maketitle

\section{\label{sec:level1}INTRODUCTION}

In the 21st century, as we navigate through an era characterized by an unprecedented rate of scientific and technological advancements and the unfolding of the fourth industrial revolution, humanity is confronted with the enormous challenge of effectively processing, utilizing, and most importantly, disseminating knowledge and skills through education. Contemporary society mandates that young graduates possess a diverse array of knowledge and skills \cite{marope_future_2017,ng_exploring_2019}, which are not only comprehensive but also directly applicable to their chosen professions. From the perspective of STEM integration in education, we speak of a transdisciplinary or even neo-disciplinary approach \cite{ng_exploring_2019} where students should “fully immerse themselves in authentic, real-world problem-solving tasks and address them using and developing appropriate skills in combinations that in a real-world sense disregard the traditional silos of disciplines”. Addressing these demands has emerged as a cardinal challenge for STEM education \cite{khine_steam_2019}.

In this paper, we introduce SageMath as a promising, open, and freely accessible digital tool in the context of STEM education and explore its role in the seamless integration of natural sciences and mathematics with digital technologies. SageMath has been acknowledged for its enormous potential by the OpenDreamKit project -- a  European Research Infrastructure project (2015-2019, \url{www.opendreamkit.org}) that involved collaborations between fifteen major European universities. The primary objective of OpenDreamKit was to develop a comprehensive and freely accessible scientific and educational digital platform, comprised exclusively of open-source software, for the needs of not only researchers, scientists, and engineers, but also teachers and students. Through its versatility and powerful features, SageMath has the potential to provide innovative ways in which STEM education is delivered, fostering an environment that nurtures critical thinking, problem-solving, and innovation.

\section{WHAT IS SAGEMATH? }
SageMath, commonly abbreviated as Sage (\url{www.sagemath.org}, \cite{stein_sage_2022}), is a free, open-source mathematics software system that is based on Python and licensed under the GPL. Guided by the motto “building the car instead of reinventing the wheel,” Sage harnesses the capabilities of several hundred existing open-source software packages and libraries, e.g. Numpy, SciPy, Sympy, Matplotlib, or Maxima, and integrates them into a cohesive learning and research experience that is well suited for both novice and experienced users.  Through this integration of various independent software components into a unified framework, SageMath establishes itself as a viable and powerful alternative to well-known commercial Computer Algebra Systems (CAS) such as Maple, Mathematica, or  Matlab \cite{zimmermann_computational_2018}.

As we said, Sage is built upon Python, a widely used general-purpose programming language known for its versatility and also prominence in the scientific computing community and its use by leading organizations such as Google and NASA. The uniqueness of Sage also lies in the fact that it not only significantly amplifies but also simplifies Python’s mathematical capabilities, as the commands and notation for mathematical computations in Sage are much closer to standard mathematical notation than is the case with Python. This makes Sage more intuitive and accessible, especially for those who are already familiar with mathematical conventions, and facilitates a smoother transition into computational mathematics.

Technologically, Sage is distributed with Jupyter (\url{www.jupyter.org}, \cite{kluyver_jupyter_2016}) -- a web interactive environment in which Sage operates as a computational kernel. More detailed information on what Jupyter is and how Sage with Jupyter can be installed is described in detail our recent article \cite{gajdos_interactive_2022}. The native document format for Jupyter, known as Jupyter Notebook, can incorporate not only code and computations but also additional narrative text, multimedia elements, and visualizations. This synergy with Jupyter enhances the user experience by facilitating dynamic, interactive, and richly annotated computational documents, making SageMath a versatile tool for both educational and research pursuits in the STEM fields.

\section{SAGEMATH IN STEM EDUCATION}

\subsection{Action research in innovating the teaching of physical problems solving}

In 2019, in line with action research principles \cite{stringer_action_2020}, authors of the paper realized a significant innovation in the course "Methods of Physical Problems Solving" for bachelors in the interdisciplinary study of Physics with other science subjects (Mathematics, Biology, Chemistry, Informatics, Geography). The following objectives were established from the students' perspective, where the student is expected to acquire:

\begin{itemize}
    \item An overview of qualitative, quantitative, and experimental methods of solving physical problems,
    \item Skills in modeling a given physical problem and applying appropriate methods of solution according to the nature of the physical problem,
    \item Skills in the effective use of digital technologies on PCs, mobile devices, and tablets in solving problems.
\end{itemize}

A significant change also occurred in using teaching methods and forms. There was extensive implementation of blended and flipped learning \cite{talbert_flipped_2017,tucker_blended_2017}, which was reflected, for example, in the conditions for course assessment, in which instead of traditional credit tests, the outcomes of students' ongoing independent work on assignments and projects were now included. Finally,  a new, innovative syllabus was developed, as depicted in Table 1 (on the left), with an emphasis on physical modeling and visualization, strongly augmented by digital technologies. This approach leverages students’ prior knowledge and skills obtained from both general and theoretical courses in STEM disciplines. Sage was adopted as the unifying and versatile digital tool. In the context of modeling as a key component and competence of instruction \cite{weber_benefit_2020}, Sage’s utility spans several areas (Table 1, on the right): (1) source of open scientific data, (2) numerical and symbolic computations, (3) data processing, (4) visualizations, (5) content presentation, and (6) programming.

One of the basic sources of physical problems that we address in this course are tasks and problems from physics competitions at the primary and secondary school levels, specifically, the Physics Olympiad (PhO) and the Young Physicists’ Tournament (YPT). The competitions offer tasks of varying difficulty levels, from interesting tasks suitable for beginner-level students (cat. F of PhO -- 7th grade) to highly complex, challenging physics problems in the YPT. 

\begin{table}
\caption{\label{tab:table1}Brief outline of our course Methods of Physical Problems Solving.}
\begin{ruledtabular}
\begin{tabular}{l l}
\rule{0pt}{1\normalbaselineskip}
Week/Theme & SageMath with Jupyter applications \\[3pt]
\hline
\rule{0pt}{1.2\normalbaselineskip}
\textbf{Introduction to the subject} & \\
1. Overview of approaches, methods and means, & \\
sources of physical problems, competitions & \\[6pt]
\textbf{Qualitative approaches} & \\
2. Simple thought modeling and Fermi estimates & scientific data source\\
3. Dimensional analysis, scaling & \\
4. Application of symmetry and conservation laws & \\
5. Graphical methods & powerful graphical calculator \\[6pt]
\textbf{Experiment and digital technologies} & \\
6. Animations and simple simulations (Geogebra, Phet, \& Workbench, Physlets) & multimedia presentations\\
7. Video analysis (Tracker), iconographic modeling (VnR, Coach) & data processing\\
8. Computer-aided, remote and virtual experiments (PC, tablet, mobile) & interactive
visualizations \& simulations
 \\[6pt]
\textbf{Quantitative approaches} & \\
9. Models in the form of differential equations - computer modeling (Sage, Jupyter) & modeling\\
10. Symbolic and numerical solutions (Sage, Jupyter) & symbolic \& numerical
solutions
\\[6pt]
\textbf{More advanced approaches }& \\
11. Qualitative approach through the theory of dynamical systems & programming
(Python, R, Julia)
\\
12. Variational approaches (Lagrange, Hamilton) & \\
13. 2D and 3D visualization and verification of solutions using a computer (Sage, VPython) & advanced interactive
visualizations
\\[3pt]
\end{tabular}
\end{ruledtabular}
\end{table}

In the following two sections, we clearly and specifically demonstrate the capabilities and features of Sage in solving two specific problems from these competitions. However, these features and capabilities can be applied across the STEM disciplines, not only in physics, but primarily in mathematics, as well as in chemistry or biology. 

\subsection{Illustrative example 1 -- The Physics Olympiad (PhO)}
In the first illustrative example from the Physics Olympiad (PhO), we present Sage as an open scientific data source, as well as a symbolic, numerical, and visualization tool. This problem is from the 63rd edition of the Slovak Physics Olympiad (regional round, category C -- 2nd year of high school) from the academic year 2021/2022 (authors: Bystrický \& Konrád). The problem statement dealing with thermodynamics, consisting of parts a) to d), can be seen in Fig. 1 (on the right). To solve the problem, Sage (Fig. 1,  \texttt{cells [1]-[3]}) provides the necessary physical constants through the Python library \texttt{scipy.constants}, such as the molar gas constant, Avogadro's number, or the value of 0°C in Kelvin. In addition, libraries \texttt{mendeleev} and \texttt{molmass} give us further chemical constants for incorporating them into the calculations. For example (\texttt{cell [4]}), we can obtain the atomic mass of nitrogen, the number of protons and neutrons in nitrogen, or the molar mass of nitrogen gas as the chemical compound N2. The same can be done for any other compound by specifying its chemical formula. An intriguing feature is also the ability to retrieve additional valuable information regarding the discovery of a chemical element or the applications of its compounds (Fig. 1,  \texttt{cells [5],[6]}).

It is important to note that the \texttt{scipy.constants }library retrieves data directly from the reputable open database of physical constants and units, CODATA2018, managed by the US National Institute of Standards and Technology (NIST, \url{https://www.nist.gov/}). Without the need for manual searching and copying of constant values, which could be prone to human error, we also have access to other valuable open scientific sources via Sage.

\begin{figure}
    \centering
    \includegraphics[width=0.75\linewidth]{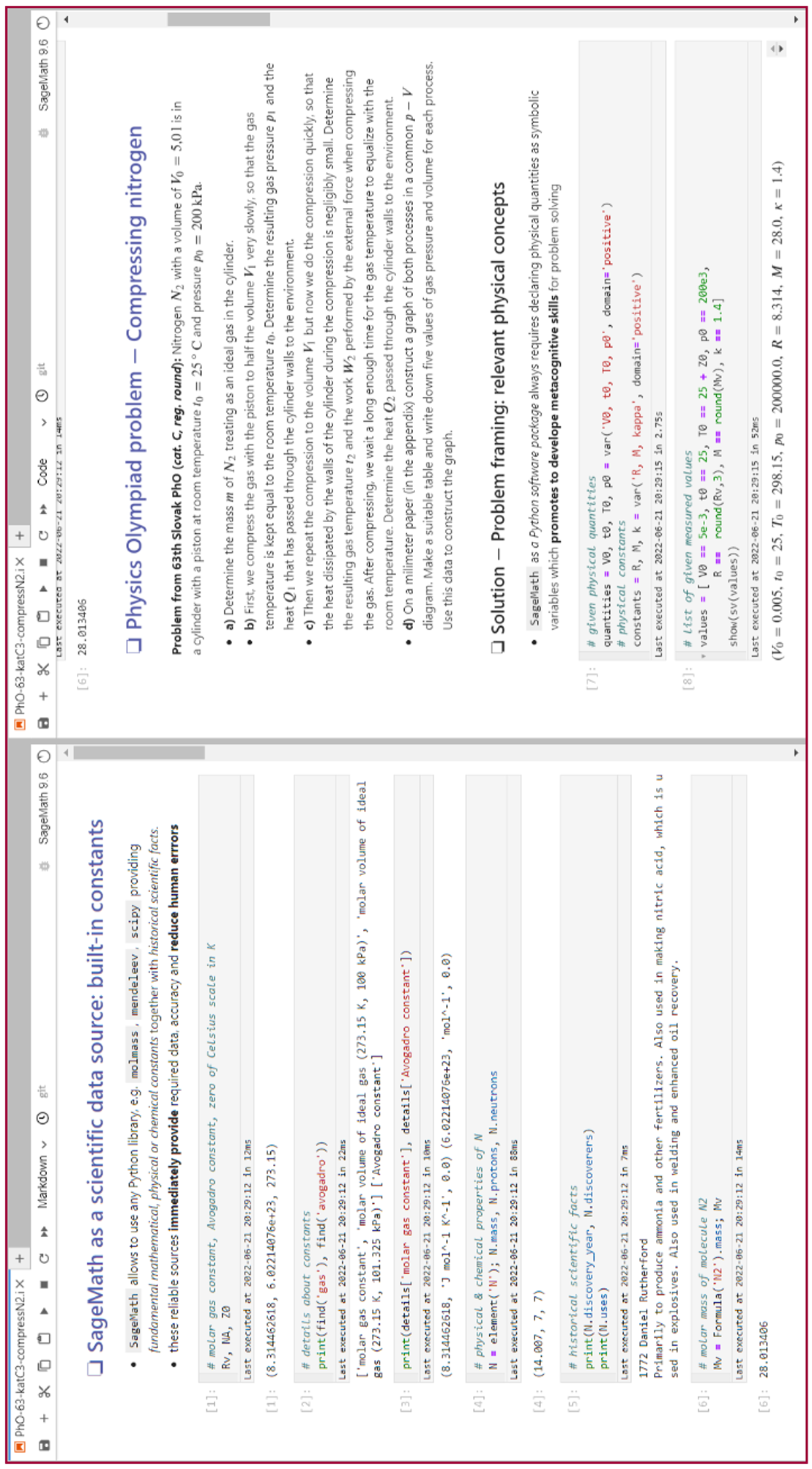}
    \caption{Screenshots of the Sage use in solving a problem from 63th Slovak Physics Olympiad, category C.}
    \label{fig:1}
\end{figure}

 \begin{figure}
    \centering
    \includegraphics[width=0.75\linewidth]{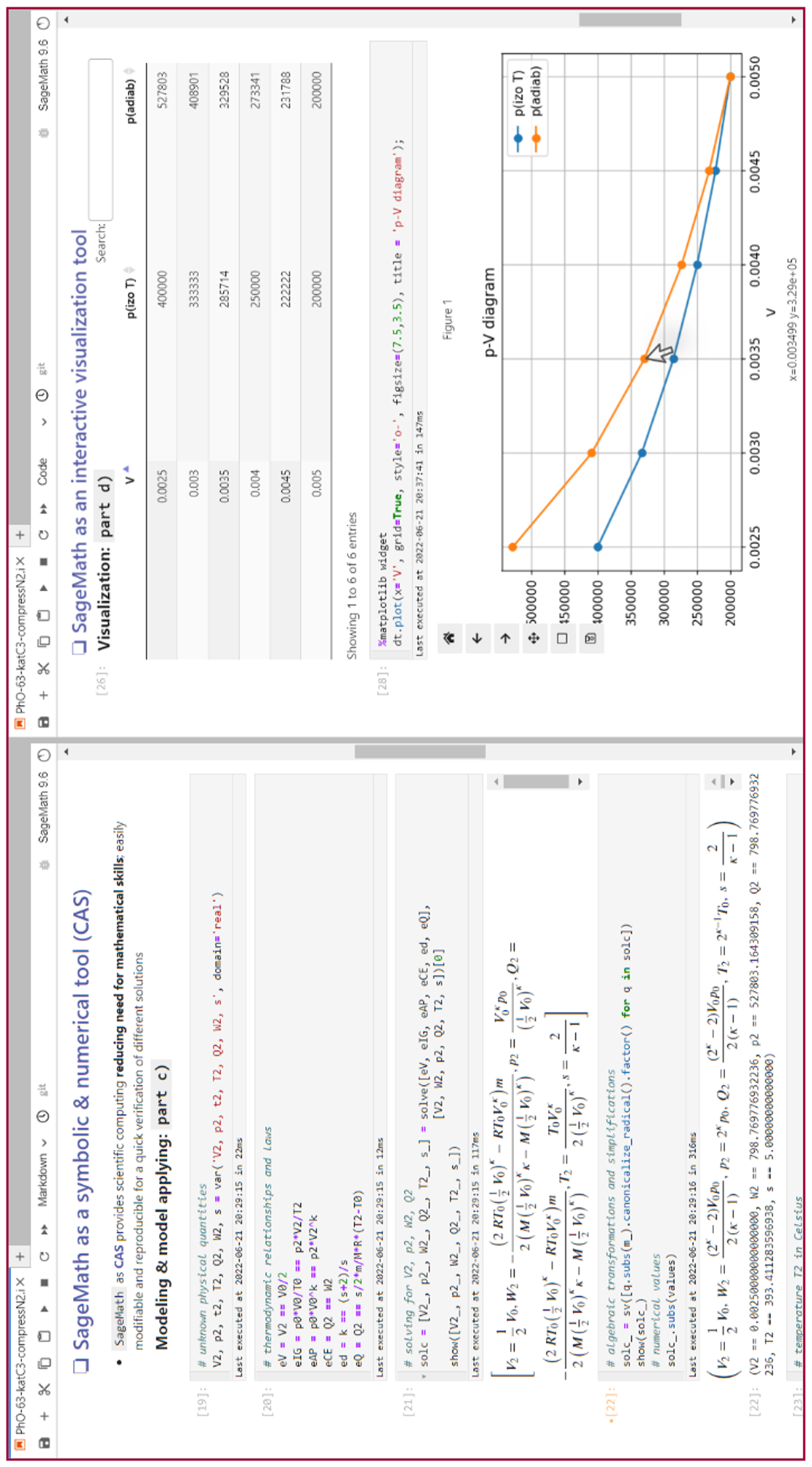}
    \caption{Screenshots of the Sage use in solving a problem from 63th Slovak Physics Olympiad, category C.}
    \label{fig:2}
\end{figure}

As examples, we mention (1) the chemical database PubChem (\url{https://pubchem.ncbi.nlm.nih.gov/}, pubchempy library), (2) the US National Institutes of Health biological database (\url{https://www.ncbi.nlm.nih.gov/}, biopython), and (3) the Materials Project database of material properties (\url{https://materialsproject.org/}, pymatgen). Moreover, SageMath offers a wide range of built-in mathematical constants and functions that can be computed with arbitrary precision or handled in symbolic computations.

Using variables and constants in Sage is the same as in Python. However, unknown physical quantities must be declared as symbolic variables, which is not a disadvantage, but rather the opposite. It promotes the development of metacognitive skills, as students need to be properly aware of relevant physical concepts which are represented by known and unknown quantities in the problem solving (see Fig. 1,  \texttt{cells [7],[8]}, and Fig. 2,  \texttt{cell [19]}). In Fig.2, we can see the solution to part c) of the problem itself. First in a general form, where in  \texttt{cell [20]} we see the physical laws specified as a system of equations that need to be solved. Using the solve command, we can easily obtain the general analytic solution to this system of non-linear equations ( \texttt{cell [21]}). 

When displaying mathematical relationships and formulas, Sage uses the LaTeX typesetting system (\url{https://www.latex-project.org/}) through the show command, which makes the output look like standard mathematics in textbooks, contributing to easy readability and navigation. By using the subs command, we can instantly get numerical solutions by substitutions of the values obtained from previous tasks. If we check the student’ solution, we can easily use the “incorrect” input values of a hypothetical student who made a numerical error into the given solution.

Another valuable Sage feature is the graphical visualization of results, which is required in the subsequent task d). In such a case, we can use the pandas library. Interestingly, the table of values for the graph can have an interactive form, where, similarly to Excel, we can sort the rows in ascending or descending order based on the values in a selected column. You can easily search for specific rows in the table, and only those rows that match the search parameters are displayed. It is also possible to display results in the form of an interactive graph using the matplotlib library. In such a case, the user can zoom in and out, move the displayed area, reset the graph display settings, copy the current displayed area to the clipboard, and also explore the graph by moving the cursor over it, displaying the corresponding x and y coordinates.

Interactive Sage solutions of PhO problems developed in this way were used in a real practice for correcting tasks in a given ongoing competition. Pilot testing in last three years showed that during corrections, the correction time decreased to about 1/3, but what we value the most is that the students’ solutions were certainly corrected more objectively and fairly, as we did not penalize a student multiple times for a single mistake.

\subsection{Illustrative example 2 -- The Young Physicist Tournament (YPT)}
The 2nd example from the 34th Young Physicists' Tournament will showcase Sage as a multimedia, modeling, and programming tool. This illustrative example concerns Problem 12, which deals with the Wilberforce pendulum, the spring-mass system that consists of a mass m with an adjustable moment of inertia I suspended from a helical spring (Fig. 3, top left). In Sage, when introducing this problem, multimedia can be utilized. For example, videos from YouTube or a custom video demonstrating the motion of the Wilberforce pendulum can be embedded. Alternatively, it is possible to insert necessary schematic illustrations or sketches to enhance the solution comprehensibility.

From a physics perspective, this time the problem belongs to the very challenging physical problems \cite{uy_wilberforce_2021}, whose effective modeling require advanced modeling methods. In the case of the Wilberforce pendulum it is the Lagrange variational approach. Our undergraduate students have experience mainly with classical mechanics in the sense of Newton's approach, in which they analyze the problem in terms of forces and operate with vectors. This is the traditional and well-learned approach, and it represents a starting point in analyzing the physical problems. 

Later, when faced with the Lagrange approach within a theoretical course of analytical mechanics, a common misconception often arises that such an approach is disproportionately more demanding and even unhelpful. In fact, when dealing with such non-trivial problems, it is Newton's approach that is more difficult, time consuming and therefore practically unusable. The Lagrange approach is based on the analysis of the problem in terms of energies scalar quantities. The advantage of this approach is the invariance of Lagrange equations under point transformations, which eliminates the consideration of whether the system is inertial or non-inertial, and the option of choosing an ideal coordinate system or a set of generalized coordinates in which the analysis is more natural and reasonable.

From a didactic viewpoint, we have to note that there is a fairly intuitive way to introduce and show the equivalence and mutual analogy of the Lagrange and Newton's approaches \cite{levi_classical_2014}. Our experience has shown us that this method is also suitable for talented high school students who participate in the Young Physicists' Tournament (YPT) and have a grasp of elementary singe variable calculus.
 
\begin{figure}
    \centering
    \includegraphics[width=0.75\linewidth]{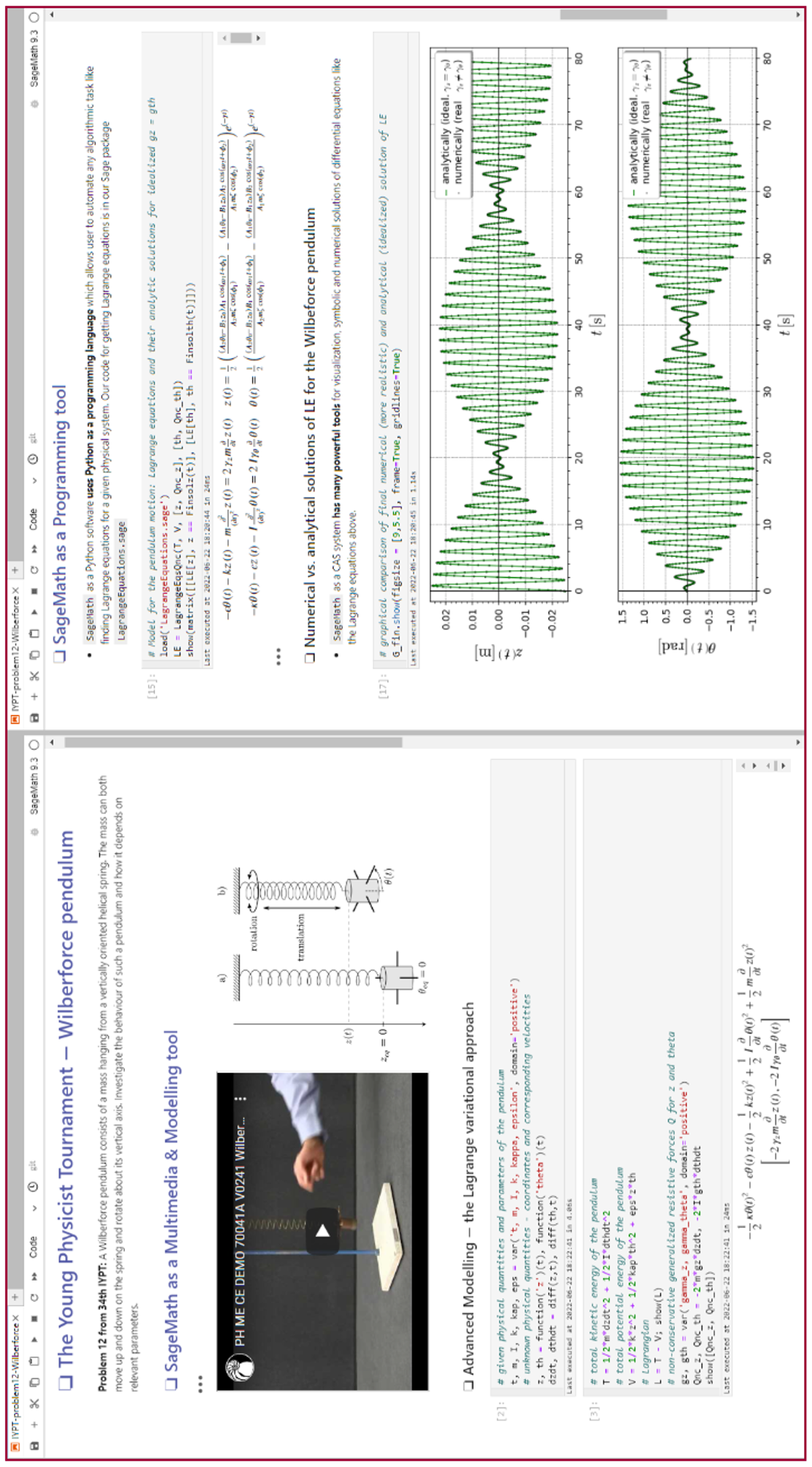}
    \caption{Screenshots of the Sage use in solving a problem from 34th The Young Physicist Tournament. }
    \label{fig:3}
\end{figure}

On the other hand, we believe that students' opinion about the higher difficulty of the Lagrange approach stems from the demanding mathematical apparatus with which this approach works. This is where Sage comes in handy with capabilities in symbolic and numerical calculations. Students need to be able to physically describe the problem using classical energy concepts and construct the Lagrangian -- the difference between the total kinetic and total potential energy of the system. The mathematical manipulations, which are involved in formulating the Lagrange equations, can be left to Sage. So, this step can be automated after introducing given quantities in Sage (see Fig. 3,  \texttt{cells [2],[3]}) using a suitably created Python function (see Fig. 3, library LagrangeEquations.sage,  \texttt{cell [15]} and its output). The coding of such a function requires only basic and novice knowledge of Python algorithmic structures. 

Furthermore, Sage can provide symbolically an analytical solution or the ability to apply the necessary substitutions and make appropriate adjustments to the expressions that will help to arrive at the desired form of solution (see Fig. 3,  \texttt{cell [15]}). The advantage is that this eliminates human-induced errors and the ability to easily edit any of the results at any stage of the solution. In the case of the Wilberforce pendulum, a meaningful analytical solution of the motion equations can only be obtained under certain specific conditions. But in the case of using Sage, we do not have to limit ourselves and can find out the numerical solution for general conditions with graphical visualizations (Fig. 3,  \texttt{cell [17]}). The dotted plot represents a numerical solution that is consistent with the analytical solution displayed as a green line. Sage provides many tools for numerical solutions of differential equations. We highlight \texttt{desolve\_odeint}, which is extremely fast in Sage due to the fact that it was written in Fortran. But nevertheless, in Sage it is possible to call it comfortably like any Python function or command.

\section{Conclusion}
Society, industry, and stakeholders are increasingly demanding that young graduates possess a broad range of knowledge and skills that can be effectively applied in practice. In response to this, the STEM education approach has been established, emphasizing transdisciplinarity or even neo-disciplinarity and connecting theoretical knowledge with real-world phenomena. From our experience spanning over 5 years, Sage (with Jupyter), emerges as a universal, interactive, and easy-to-use scientific computing platform that can be successfully and reciprocally applied across all STEM disciplines \cite{gajdos_interactive_2022,garfinkel_modeling_2017,kostur_problems_2019}, as well as in research. For instance, we have successfully used Sage for exponentially fast, highly accurate, and very reliable calculations of the gamma difference distribution (GDD) --- the difference $X \equiv X_1 - X_2$ of two independent random variables X1,  X2 having a gamma distribution \cite{hancova_practical_2022}. Furthermore, Sage opens the gates for teaching, learning, and developing advanced modeling competencies in STEM through engagement with authentic, real-world problem-solving tasks, adding a new dimension to STEM education. It's also important to note that Sage aligns well with modern innovative pedagogical approaches such as inquiry-based learning \cite{constantinou_what_2018}, project-based learning, and blended learning \cite{talbert_flipped_2017,tucker_blended_2017}.

By employing SageMath as an open digital environment for scientific computing, both teachers and students are equipped with more than just an educational tool. They receive a genuine research instrument used by scientists for interactive visualization, modeling, programming, and tackling authentic, complex interdisciplinary problems. This naturally boosts motivation to study science, in alignment with the core mission of STEM education. Moreover, students who develop digital skills through SageMath are readily equipped to engage in real scientific tasks as student research assistants or later as Ph.D. students (as one of the authors -- DB). From a research perspective, this essential interconnection leads to much broader opportunities to test, inspect, and understand the advantages and limitations of Sage's computational tools and algorithms in the overlapping context of education and research.

\begin{acknowledgments}
This work is supported by the Slovak Research and Development Agency under the Contract no. APVV-21-0216 and APVV-21-0369.
\end{acknowledgments}

\bibliography{ReferencesDB}

\begin{thebibliography}{17}%
\makeatletter
\providecommand \@ifxundefined [1]{%
 \@ifx{#1\undefined}
}%
\providecommand \@ifnum [1]{%
 \ifnum #1\expandafter \@firstoftwo
 \else \expandafter \@secondoftwo
 \fi
}%
\providecommand \@ifx [1]{%
 \ifx #1\expandafter \@firstoftwo
 \else \expandafter \@secondoftwo
 \fi
}%
\providecommand \natexlab [1]{#1}%
\providecommand \enquote  [1]{``#1''}%
\providecommand \bibnamefont  [1]{#1}%
\providecommand \bibfnamefont [1]{#1}%
\providecommand \citenamefont [1]{#1}%
\providecommand \href@noop [0]{\@secondoftwo}%
\providecommand \href [0]{\begingroup \@sanitize@url \@href}%
\providecommand \@href[1]{\@@startlink{#1}\@@href}%
\providecommand \@@href[1]{\endgroup#1\@@endlink}%
\providecommand \@sanitize@url [0]{\catcode `\\12\catcode `\$12\catcode
  `\&12\catcode `\#12\catcode `\^12\catcode `\_12\catcode `\%12\relax}%
\providecommand \@@startlink[1]{}%
\providecommand \@@endlink[0]{}%
\providecommand \url  [0]{\begingroup\@sanitize@url \@url }%
\providecommand \@url [1]{\endgroup\@href {#1}{\urlprefix }}%
\providecommand \urlprefix  [0]{URL }%
\providecommand \Eprint [0]{\href }%
\providecommand \doibase [0]{http://dx.doi.org/}%
\providecommand \selectlanguage [0]{\@gobble}%
\providecommand \bibinfo  [0]{\@secondoftwo}%
\providecommand \bibfield  [0]{\@secondoftwo}%
\providecommand \translation [1]{[#1]}%
\providecommand \BibitemOpen [0]{}%
\providecommand \bibitemStop [0]{}%
\providecommand \bibitemNoStop [0]{.\EOS\space}%
\providecommand \EOS [0]{\spacefactor3000\relax}%
\providecommand \BibitemShut  [1]{\csname bibitem#1\endcsname}%
\let\auto@bib@innerbib\@empty
\bibitem [{\citenamefont {Marope}(2017)}]{marope_future_2017}%
  \BibitemOpen
  \bibfield  {author} {\bibinfo {author} {\bibfnamefont {M.}~\bibnamefont
  {Marope}},\ }\bibfield  {title} {\enquote {\bibinfo {title} {Future
  {Competences} for {Future} {Generations}},}\ }\href
  {https://unesdoc.unesco.org/ark:/48223/pf0000366753} {\bibfield  {journal}
  {\bibinfo  {journal} {IBE in Focus, UNESCO International Bureau of
  Education}\ ,\ \bibinfo {pages} {81--88}} (\bibinfo {year}
  {2017})}\BibitemShut {NoStop}%
\bibitem [{\citenamefont {Ng}(2019)}]{ng_exploring_2019}%
  \BibitemOpen
  \bibfield  {author} {\bibinfo {author} {\bibfnamefont {S.~B.}\ \bibnamefont
  {Ng}},\ }\bibfield  {title} {\enquote {\bibinfo {title} {Exploring {STEM}
  {Competences} for the 21st {Century}},}\ }\href {\doibase IBE/2019/WP/CD/30
  REV} {\bibfield  {journal} {\bibinfo  {journal} {In-Progress Reflection,
  UNESCO International Bureau of Education}\ }\bibinfo {series} {Current and
  {Critical} {Issues} in {Curriculum}, {Learning} and {Assessment}},\ \bibinfo
  {pages} {53} (\bibinfo {year} {2019})}\BibitemShut {NoStop}%
\bibitem [{\citenamefont {Khine}\ and\ \citenamefont
  {Areepattamannil}(2019)}]{khine_steam_2019}%
  \BibitemOpen
  \bibfield  {author} {\bibinfo {author} {\bibfnamefont {M.~S.}\ \bibnamefont
  {Khine}}\ and\ \bibinfo {author} {\bibfnamefont {S.}~\bibnamefont
  {Areepattamannil}},\ }\href@noop {} {\emph {\bibinfo {title} {{STEAM}
  {Education}: {Theory} and {Practice}}}}\ (\bibinfo  {publisher} {Springer},\
  \bibinfo {address} {Cham},\ \bibinfo {year} {2019})\BibitemShut {NoStop}%
\bibitem [{\citenamefont {Stein}\ and\ \citenamefont
  {{others}}(2022)}]{stein_sage_2022}%
  \BibitemOpen
  \bibfield  {author} {\bibinfo {author} {\bibfnamefont {W.~A.}\ \bibnamefont
  {Stein}}\ and\ \bibinfo {author} {\bibnamefont {{others}}},\ }\href
  {http://www.sagemath.org} {\enquote {\bibinfo {title} {Sage {Mathematics}
  {Software} - {SageMath}},}\ } (\bibinfo {year} {2022})\BibitemShut {NoStop}%
\bibitem [{\citenamefont {Zimmermann}\ \emph {et~al.}(2018)\citenamefont
  {Zimmermann}, \citenamefont {Casamayou}, \citenamefont {Cohen}, \citenamefont
  {Connan}, \citenamefont {Dumont}, \citenamefont {Fousse}, \citenamefont
  {Maltey}, \citenamefont {Meulien}, \citenamefont {Mezzarobba}, \citenamefont
  {Pernet}, \citenamefont {Thi{\'e}ry}, \citenamefont {Bray}, \citenamefont
  {Cremona}, \citenamefont {Forets}, \citenamefont {Ghitza},\ and\
  \citenamefont {Thomas}}]{zimmermann_computational_2018}%
  \BibitemOpen
  \bibfield  {author} {\bibinfo {author} {\bibfnamefont {P.}~\bibnamefont
  {Zimmermann}}, \bibinfo {author} {\bibfnamefont {A.}~\bibnamefont
  {Casamayou}}, \bibinfo {author} {\bibfnamefont {N.}~\bibnamefont {Cohen}},
  \bibinfo {author} {\bibfnamefont {G.}~\bibnamefont {Connan}}, \bibinfo
  {author} {\bibfnamefont {T.}~\bibnamefont {Dumont}}, \bibinfo {author}
  {\bibfnamefont {L.}~\bibnamefont {Fousse}}, \bibinfo {author} {\bibfnamefont
  {F.}~\bibnamefont {Maltey}}, \bibinfo {author} {\bibfnamefont
  {M.}~\bibnamefont {Meulien}}, \bibinfo {author} {\bibfnamefont
  {M.}~\bibnamefont {Mezzarobba}}, \bibinfo {author} {\bibfnamefont
  {C.}~\bibnamefont {Pernet}}, \bibinfo {author} {\bibfnamefont {N.~M.}\
  \bibnamefont {Thi{\'e}ry}}, \bibinfo {author} {\bibfnamefont
  {E.}~\bibnamefont {Bray}}, \bibinfo {author} {\bibfnamefont {J.}~\bibnamefont
  {Cremona}}, \bibinfo {author} {\bibfnamefont {M.}~\bibnamefont {Forets}},
  \bibinfo {author} {\bibfnamefont {A.}~\bibnamefont {Ghitza}}, \ and\ \bibinfo
  {author} {\bibfnamefont {H.}~\bibnamefont {Thomas}},\ }\href@noop {} {\emph
  {\bibinfo {title} {Computational {Mathematics} with {SageMath}}}},\ \bibinfo
  {edition} {1st}\ ed.\ (\bibinfo  {publisher} {SIAM},\ \bibinfo {address}
  {Philadelphia},\ \bibinfo {year} {2018})\BibitemShut {NoStop}%
\bibitem [{\citenamefont {Kluyver}\ \emph {et~al.}(2016)\citenamefont
  {Kluyver}, \citenamefont {Ragan-Kelley}, \citenamefont {Perez}, \citenamefont
  {Granger}, \citenamefont {Bussonnier}, \citenamefont {Frederic},
  \citenamefont {Kelley}, \citenamefont {Hamrick}, \citenamefont {Grout},
  \citenamefont {Corlay}, \citenamefont {Ivanov}, \citenamefont {Avila},
  \citenamefont {Abdalla},\ and\ \citenamefont
  {Willing}}]{kluyver_jupyter_2016}%
  \BibitemOpen
  \bibfield  {author} {\bibinfo {author} {\bibfnamefont {T.}~\bibnamefont
  {Kluyver}}, \bibinfo {author} {\bibfnamefont {B.}~\bibnamefont
  {Ragan-Kelley}}, \bibinfo {author} {\bibfnamefont {F.}~\bibnamefont {Perez}},
  \bibinfo {author} {\bibfnamefont {B.}~\bibnamefont {Granger}}, \bibinfo
  {author} {\bibfnamefont {M.}~\bibnamefont {Bussonnier}}, \bibinfo {author}
  {\bibfnamefont {J.}~\bibnamefont {Frederic}}, \bibinfo {author}
  {\bibfnamefont {K.}~\bibnamefont {Kelley}}, \bibinfo {author} {\bibfnamefont
  {J.}~\bibnamefont {Hamrick}}, \bibinfo {author} {\bibfnamefont
  {J.}~\bibnamefont {Grout}}, \bibinfo {author} {\bibfnamefont
  {S.}~\bibnamefont {Corlay}}, \bibinfo {author} {\bibfnamefont
  {P.}~\bibnamefont {Ivanov}}, \bibinfo {author} {\bibfnamefont
  {D.}~\bibnamefont {Avila}}, \bibinfo {author} {\bibfnamefont
  {S.}~\bibnamefont {Abdalla}}, \ and\ \bibinfo {author} {\bibfnamefont
  {C.}~\bibnamefont {Willing}},\ }\bibfield  {title} {\enquote {\bibinfo
  {title} {Jupyter {Notebooks}-a publishing format for reproducible
  computational workflows},}\ }in\ \href@noop {} {\emph {\bibinfo {booktitle}
  {Positioning and {Power} in {Academic} {Publishing}: {Players}, {Agents} and
  {Agendas}. {Proceedings} of the 20th {International} {Conference} on
  {Electronic} {Publishing}.}}},\ \bibinfo {editor} {edited by\ \bibinfo
  {editor} {\bibfnamefont {F.}~\bibnamefont {Loizides}}\ and\ \bibinfo {editor}
  {\bibfnamefont {B.}~\bibnamefont {Schmidt}}}\ (\bibinfo  {publisher} {Ios
  Press},\ \bibinfo {address} {Amsterdam},\ \bibinfo {year} {2016})\ pp.\
  \bibinfo {pages} {87 -- 90}\BibitemShut {NoStop}%
\bibitem [{\citenamefont {Gajdo{\v s}}, \citenamefont {Han{\v c}},\ and\
  \citenamefont {Han{\v c}ov{\'a}}(2022)}]{gajdos_interactive_2022}%
  \BibitemOpen
  \bibfield  {author} {\bibinfo {author} {\bibfnamefont {A.}~\bibnamefont
  {Gajdo{\v s}}}, \bibinfo {author} {\bibfnamefont {J.}~\bibnamefont {Han{\v
  c}}}, \ and\ \bibinfo {author} {\bibfnamefont {M.}~\bibnamefont {Han{\v
  c}ov{\'a}}},\ }\bibfield  {title} {\enquote {\bibinfo {title} {Interactive
  {Jupyter} {Notebooks} with {SageMath} in {Number} {Theory}, {Algebra},
  {Calculus}, and {Numerical} {Methods}},}\ }in\ \href
  {https://ieeexplore.ieee.org/document/9974868} {\emph {\bibinfo {booktitle}
  {Proceedings of {ICETA} 2022 ({October} 20-21, {Star{\'y}} {Smokovec},
  2022)}}},\ \bibinfo {series and number} {\bibinfo {number} {CFP2238M-USB}},\
  \bibinfo {editor} {edited by\ \bibinfo {editor} {\bibfnamefont
  {F.}~\bibnamefont {Jakab}}}\ (\bibinfo  {publisher} {IEEE},\ \bibinfo
  {address} {Danvers},\ \bibinfo {year} {2022})\ pp.\ \bibinfo {pages}
  {166--171}\BibitemShut {NoStop}%
\bibitem [{\citenamefont {Stringer}\ and\ \citenamefont
  {Arag{\'o}n}(2020)}]{stringer_action_2020}%
  \BibitemOpen
  \bibfield  {author} {\bibinfo {author} {\bibfnamefont {E.~T.}\ \bibnamefont
  {Stringer}}\ and\ \bibinfo {author} {\bibfnamefont {A.~O.}\ \bibnamefont
  {Arag{\'o}n}},\ }\href@noop {} {\emph {\bibinfo {title} {Action
  {Research}}}},\ \bibinfo {number} {5}\ (\bibinfo  {publisher} {SAGE},\
  \bibinfo {address} {Thousand Oaks},\ \bibinfo {year} {2020})\BibitemShut
  {NoStop}%
\bibitem [{\citenamefont {Talbert}\ and\ \citenamefont
  {Bergmann}(2017)}]{talbert_flipped_2017}%
  \BibitemOpen
  \bibfield  {author} {\bibinfo {author} {\bibfnamefont {R.}~\bibnamefont
  {Talbert}}\ and\ \bibinfo {author} {\bibfnamefont {J.}~\bibnamefont
  {Bergmann}},\ }\href@noop {} {\emph {\bibinfo {title} {Flipped {Learning}:
  {A} {Guide} for {Higher} {Education} {Faculty}}}}\ (\bibinfo  {publisher}
  {Stylus Publishing},\ \bibinfo {address} {Sterling, Virginia},\ \bibinfo
  {year} {2017})\BibitemShut {NoStop}%
\bibitem [{\citenamefont {Tucker}, \citenamefont {Wycoff},\ and\ \citenamefont
  {Green}(2017)}]{tucker_blended_2017}%
  \BibitemOpen
  \bibfield  {author} {\bibinfo {author} {\bibfnamefont {C.~R.}\ \bibnamefont
  {Tucker}}, \bibinfo {author} {\bibfnamefont {T.}~\bibnamefont {Wycoff}}, \
  and\ \bibinfo {author} {\bibfnamefont {J.~T.}\ \bibnamefont {Green}},\
  }\href@noop {} {\emph {\bibinfo {title} {Blended {Learning} in {Action}: {A}
  {Practical} {Guide} {Toward} {Sustainable} {Change}}}}\ (\bibinfo
  {publisher} {Corwin Press},\ \bibinfo {address} {Thousand Oaks},\ \bibinfo
  {year} {2017})\BibitemShut {NoStop}%
\bibitem [{\citenamefont {Weber}\ and\ \citenamefont
  {Wilhelm}(2020)}]{weber_benefit_2020}%
  \BibitemOpen
  \bibfield  {author} {\bibinfo {author} {\bibfnamefont {J.}~\bibnamefont
  {Weber}}\ and\ \bibinfo {author} {\bibfnamefont {T.}~\bibnamefont
  {Wilhelm}},\ }\bibfield  {title} {\enquote {\bibinfo {title} {The benefit of
  computational modelling in physics teaching: a historical overview},}\ }\href
  {https://doi.org/10.1088/1361-6404/ab7a7f} {\bibfield  {journal} {\bibinfo
  {journal} {Eur. J. Phys.}\ }\textbf {\bibinfo {volume} {41}},\ \bibinfo
  {pages} {034003} (\bibinfo {year} {2020})}\BibitemShut {NoStop}%
\bibitem [{\citenamefont {Uy}\ \emph {et~al.}(2021)\citenamefont {Uy},
  \citenamefont {Yuan}, \citenamefont {Chai},\ and\ \citenamefont
  {Khor}}]{uy_wilberforce_2021}%
  \BibitemOpen
  \bibfield  {author} {\bibinfo {author} {\bibfnamefont {R.~F.}\ \bibnamefont
  {Uy}}, \bibinfo {author} {\bibfnamefont {C.}~\bibnamefont {Yuan}}, \bibinfo
  {author} {\bibfnamefont {Z.}~\bibnamefont {Chai}}, \ and\ \bibinfo {author}
  {\bibfnamefont {J.}~\bibnamefont {Khor}},\ }\bibfield  {title} {\enquote
  {\bibinfo {title} {Wilberforce pendulum: modelling linearly damped coupled
  oscillations of a spring-mass system},}\ }\href
  {https://doi.org/10.1088/1361-6404/ac3ac8} {\bibfield  {journal} {\bibinfo
  {journal} {Eur. J. Phys.}\ }\textbf {\bibinfo {volume} {43}},\ \bibinfo
  {pages} {015011} (\bibinfo {year} {2021})}\BibitemShut {NoStop}%
\bibitem [{\citenamefont {Levi}(2014)}]{levi_classical_2014}%
  \BibitemOpen
  \bibfield  {author} {\bibinfo {author} {\bibfnamefont {M.}~\bibnamefont
  {Levi}},\ }\href@noop {} {\emph {\bibinfo {title} {Classical {Mechanics}
  {With} {Calculus} of {Variations} and {Optimal} {Control}: {An} {Intuitive}
  {Introduction}}}},\ \bibinfo {edition} {1st}\ ed.\ (\bibinfo  {publisher}
  {American Mathematical Society},\ \bibinfo {address} {Providence},\ \bibinfo
  {year} {2014})\BibitemShut {NoStop}%
\bibitem [{\citenamefont {Garfinkel}, \citenamefont {Shevtsov},\ and\
  \citenamefont {Guo}(2017)}]{garfinkel_modeling_2017}%
  \BibitemOpen
  \bibfield  {author} {\bibinfo {author} {\bibfnamefont {A.}~\bibnamefont
  {Garfinkel}}, \bibinfo {author} {\bibfnamefont {J.}~\bibnamefont {Shevtsov}},
  \ and\ \bibinfo {author} {\bibfnamefont {Y.}~\bibnamefont {Guo}},\
  }\href@noop {} {\emph {\bibinfo {title} {Modeling {Life}: {The} {Mathematics}
  of {Biological} {Systems}}}}\ (\bibinfo  {publisher} {Springer International
  Publishing},\ \bibinfo {address} {New York},\ \bibinfo {year}
  {2017})\BibitemShut {NoStop}%
\bibitem [{\citenamefont {Kostur}, \citenamefont {{\L }uczka},\ and\
  \citenamefont {Machura}(2019)}]{kostur_problems_2019}%
  \BibitemOpen
  \bibfield  {author} {\bibinfo {author} {\bibfnamefont {M.}~\bibnamefont
  {Kostur}}, \bibinfo {author} {\bibfnamefont {J.}~\bibnamefont {{\L }uczka}},
  \ and\ \bibinfo {author} {\bibfnamefont {{\L }.}~\bibnamefont {Machura}},\
  }\href {https://github.com/marcinofulus} {\emph {\bibinfo {title} {Problems
  in {Physics} with {Sage}: {Interactive} textbooks}}}\ (\bibinfo  {publisher}
  {University of Silesia},\ \bibinfo {address} {Katowice},\ \bibinfo {year}
  {2019})\BibitemShut {NoStop}%
\bibitem [{\citenamefont {Han{\v c}ov{\'a}}, \citenamefont {Gajdo{\v s}},\ and\
  \citenamefont {Han{\v c}}(2022)}]{hancova_practical_2022}%
  \BibitemOpen
  \bibfield  {author} {\bibinfo {author} {\bibfnamefont {M.}~\bibnamefont
  {Han{\v c}ov{\'a}}}, \bibinfo {author} {\bibfnamefont {A.}~\bibnamefont
  {Gajdo{\v s}}}, \ and\ \bibinfo {author} {\bibfnamefont {J.}~\bibnamefont
  {Han{\v c}}},\ }\bibfield  {title} {\enquote {\bibinfo {title} {A practical,
  effective calculation of gamma difference distributions with open data
  science tools},}\ }\href
  {https://www.tandfonline.com/doi/full/10.1080/00949655.2021.2023873}
  {\bibfield  {journal} {\bibinfo  {journal} {J. Stat. Comput. Simul.}\
  }\textbf {\bibinfo {volume} {92}},\ \bibinfo {pages} {2205--2232} (\bibinfo
  {year} {2022})}\BibitemShut {NoStop}%
\bibitem [{\citenamefont {Constantinou}, \citenamefont {Tsivitanidou},\ and\
  \citenamefont {Rybska}(2018)}]{constantinou_what_2018}%
  \BibitemOpen
  \bibfield  {author} {\bibinfo {author} {\bibfnamefont {C.~P.}\ \bibnamefont
  {Constantinou}}, \bibinfo {author} {\bibfnamefont {O.~E.}\ \bibnamefont
  {Tsivitanidou}}, \ and\ \bibinfo {author} {\bibfnamefont {E.}~\bibnamefont
  {Rybska}},\ }\bibfield  {title} {\enquote {\bibinfo {title} {What {Is}
  {Inquiry}-{Based} {Science} {Teaching} and {Learning}?}}\ }in\ \href@noop {}
  {\emph {\bibinfo {booktitle} {Professional {Development} for
  {Inquiry}-{Based} {Science} {Teaching} and {Learning}}}},\ \bibinfo {series
  and number} {Contributions from {Science} {Education} {Research}},\ \bibinfo
  {editor} {edited by\ \bibinfo {editor} {\bibfnamefont {O.~E.}\ \bibnamefont
  {Tsivitanidou}}, \bibinfo {editor} {\bibfnamefont {P.}~\bibnamefont {Gray}},
  \bibinfo {editor} {\bibfnamefont {E.}~\bibnamefont {Rybska}}, \bibinfo
  {editor} {\bibfnamefont {L.}~\bibnamefont {Louca}}, \ and\ \bibinfo {editor}
  {\bibfnamefont {C.~P.}\ \bibnamefont {Constantinou}}}\ (\bibinfo  {publisher}
  {Springer International Publishing},\ \bibinfo {address} {Cham},\ \bibinfo
  {year} {2018})\ pp.\ \bibinfo {pages} {1--23}\BibitemShut {NoStop}%
\end{thebibliography}%

\end{document}